\documentstyle[preprint,pre,aps,psfig]{revtex}
\tightenlines
\begin{document}

\draft

% for two column  activate the line below...
%\twocolumn[\hsize\textwidth\columnwidth\hsize\csname@twocolumnfalse\endcsname
\title{Different routes to chaos via strange nonchaotic attractor in a
quasiperiodically forced system}

\author{A. Venkatesan and M. Lakshmanan}
\address{Centre for Nonlinear dynamics,\\ Department of Physics,\\
Bharathidasan University,\\Tiruchirapalli 620 024, INDIA }

\maketitle

\begin{abstract}
This paper focusses attention on the strange nonchaotic attractors (SNA) of a quasiperiodically
forced dynamical system. Several routes, including the standard ones by which the appearance of
strange nonchaotic attractors takes place, are shown to be realizable in the same
model over a two parameters ($f$-$\epsilon$) domain of the system.  In particular, the transition through  torus doubling
to chaos via SNA, torus breaking to chaos via SNA and period doubling bifurcations of
fractal torus  are demonstrated with the aid of the two 
parameter ($f$-$\epsilon$) phase diagram.
More interestingly, in order to approach the strange nonchaotic attractor, the existence of several new
bifurcations on the torus corresponding to the novel phenomenon of torus bubbling are described. Particularly,
we point out the new routes to chaos,
namely, (1) two frequency quasiperiodicity $\rightarrow $ torus doubling $\rightarrow $
torus merging followed by the gradual fractalization of torus to chaos,
(2) two frequency quasiperiodicity $\rightarrow $ torus doubling $\rightarrow$ wrinkling
$\rightarrow $ SNA $\rightarrow $ chaos $\rightarrow $ SNA $\rightarrow $ wrinkling $\rightarrow $ inverse torus doubling 
$\rightarrow $ torus $\rightarrow $ torus bubbles followed by the onset of torus breaking to chaos via SNA or 
followed by the onset of torus doubling route to chaos via SNA. The existence of the strange 
nonchaotic attractor is confirmed by calculating several characterizing quantities such as Lyapunov exponents,
winding numbers, power spectral measures and dimensions. The mechanism behind the various bifurcations are also
briefly discussed.

\end{abstract}

%\narrowtext
%\vskip1pc]
\section{ INTRODUCTION}

In nonlinear dynamical systems  strange nonchaotic attractors (SNA) are considered
as complicated structures in phase space, which is a property usually associated
with  chaotic
attractors. The pioneering work of Grebogi et al. [1]  revealed that
there are some possibilities of strange attractors in  certain types of
dynamical systems which are not chaotic.  These strange attractors are
strange in the spirit that geometrically they are strange (fractal dimensional)
objects in phase space.  On the other hand, they would not exhibit sensitivity
to initial conditions (for example, Lyapunov exponents are negative), and hence
are not chaotic.  These strange nonchaotic attractors can arise in physically
relevant situations such as quasiperiodically forced pendulum  [2-4], quantum
particles in quasiperiodic potentials [5], biological oscillators [6], Duffing
type oscillators [7-10], velocity dependent oscillators [11], electronic circuits
[12,13] and in certain maps [14-22].  Also these exotic attractors were
confirmed by an experiment consisting of a quasiperiodically forced, buckled
magnetoelastic ribbon [23], in analog simulations of a multistable potential [24]
and in a neon glow discharge experiment [25].

\par
While the existence of strange nonchaotic attractors has been firmly established,
a question that remains interesting is what are the possible routes by which they 
arise and ultimately become chaotic, and how do these attractors are born in a system (mechanism).
Several routes have been identified  in  recent times and for a few of them typical mechanisms
have also been found for the creation of SNA.  The major routes by which the SNA appear,
may be broadly classified as follows.

\par
{\bf 1. Ding et al. route [6]:} two frequency quasiperiodicity $\rightarrow$
three frequency quasiperiodicity $\rightarrow$ strange nonchaotic attractor
$\rightarrow $ chaos.

\par
{\bf 2. Kapitaniak et al. route [10]:} two frequency quasiperiodicity $\rightarrow$
strange nonchaotic attractor $\rightarrow$ three frequency quasiperiodicity
$\rightarrow$ chaos.

{\bf 3. Heagy and Hammel route [14]:}two frequency quasiperiodicity
$\rightarrow$ torus doubling $\rightarrow$ wrinkling $\rightarrow$ strange nonchaotic
attractors $\rightarrow$ chaos.

\par
{\bf 4. Feudal et al. route [17]:} two frequency quasiperiodicity $\rightarrow$wrinkling $\rightarrow$ 
strange nonchaotic attractors $\rightarrow$ chaos. 
\par
{\bf 5. Yalencinkaya and Lai route [9]:} two frequency quasiperiodicity $\rightarrow$ strange
nonchaotic attractor (on-off intermittency type attractor) $\rightarrow$
chaos.

\par
{\bf 6. Venkatesan and Lakshmanan route [11]:} two frequency quasiperiodicity $\rightarrow$
torus doubling $\rightarrow$ torus merging $\rightarrow$ wrinkling $\rightarrow$ 
strange nonchaotic attractor $\rightarrow$ chaos.

\par
{\bf 7. Kapitaniak and Chua route [13]:} two frequency quasiperiodicity
$\rightarrow$ strange nonchaotic trajectories on torus $\rightarrow$ chaos.

\par
{\bf 8. Nishikawa and Kaneko route [21]:}two frequency quasiperiodicity $\rightarrow$ 
strange nonchaotic attractors $\rightarrow$ chaos.

Different mechanisms have been identified for some of the above routes. In particular, it has been
shown that 
the birth of SNA in the Heagy and Hammel [14] route is due to the collision between a period
doubled torus and its unstable torus.  Feudal et al.[17]explained in their route that the SNA also appears 
a result of a collision of stable and unstable torus in a dense of set of points. However, Nishikawa and
Kaneko [21] discussed in their route that  the SNA emerges without interaction of stable and unstable torus.
Moreover, the loss of transverse stability of the torus can also lead to the birth of SNA, as in the case of 
the Yalencinkaya and Lai route [9]  above. For other routes as far as our knowledge goes mechanisms have
not yet been found.  

Also, most studies of strange nonchaotic attractors have focussed on their
characterization using the spectral properties[2,5], geometrical properties[6],
local divergence of trajectories[10], phase sensitivity and rational bifurcations [15-18],
and functional maps and invariant curves [20-21].

In this paper, we demonstrate the existence of at least five different routes to chaos
via  strange nonchaotic attractors in a single dynamical system namely a 
quasiperiodically forced velocity
dependent nonpolynomial oscillator system over a  two parameters ($f$-$\epsilon$) space.
To start with the birth of the  strange nonchaotic attractors  associated with
two important routes, namely, (i) torus breaking  and (ii) torus
doubling, have been studied in our model. In low dimensions, Bier and Bountis  have shown  that a dynamical system that undergoes one 
or more period 
doublings need not complete the entire infinite Feigenbaum cascade, but it may be possible to have only a finite number of period 
doublings, followed by for example  undoublings or  other bifurcations [28]. The possibility of such a novel remerging bifucation 
phenomenon of the torus doubling sequence in the quasiperiodically forced system has not yet been
reported.  Since the  system  that we 
consider possesses more than one control parameter and remains invariant under the reflection symmetry, the remergence
is likely to occur   as in the case of low dimensional systems [28,29]. To confirm such a possibility, our numerical studies show 
that  in some regions of the ($f$-$\epsilon$) parameter space,
torus doubled  orbit emerges and remerges from a single torus orbit at two different parameter values of $\epsilon$ to form 
a torus bubble. Such a remerging bifurcation  can lead to
tame the growth of the torus doubled trees and the development of the associated universal route to chaos further.
However, the nature of remerging torus doubled trees or more specifially torus bubbling, ensures the existence
of different routes for the creation of SNA when the full range of parameters are taken into account.
To illustrate these
possiblities in our system, we enumerate two new types of routes as:
(1) two frequency quasiperiodicity $\rightarrow $ torus doubling $\rightarrow $
torus merging followed by the gradual fractalization of torus to chaos [11],
(2) two frequency quasiperiodicity $\rightarrow $ torus doubling $\rightarrow$ wrinkling
$\rightarrow $ SNA $\rightarrow $ chaos $\rightarrow $ SNA $\rightarrow $ wrinkling $\rightarrow $ inverse torus doubling 
$\rightarrow $ torus $\rightarrow $ torus bubbles followed by the onset of torus breaking to chaos via SNA or 
followed by the onset of torus doubling route to chaos via strange nonchaotic attractor. 
Finally,  we also show the occurrence of period doubling
bifurcations of the destroyed torus (strange nonchaotic attractor) in our
model. 

This paper is organised as follows. Sec. 2 describes the system and the salient
features of its dynamics. Sec. 3 describes some of the  characterizing quantities 
of strange nonchaotic attractors in comparison with  chaotic attractors
such as Lyapunov exponents, winding number, power spectral analysis
and dimensions. These quantities  have been used to distinguish
between quasiperiodic, strange nonchaotic and chaotic attractors.
The birth of strange nonchaotic attractors from the transitions of the 
two frequency quasiperiodic attractors are observed in the five different routes
mentioned above  in sec. 4.  The first one is that of torus breaking to chaos via SNA. The second one is through torus 
bubbling followed by the gradual fractalization to chaos. The third route is
 the transition from torus
doubling to chaos via strange nonchaotic attractor. The next one we notice is the possibilty of
torus doubling to chaos via SNA followed by the inversely advancing type of torus. The last one we observe is  
the period doubling bifurcation of the destroyed torus.
Finally, in sec. 5 we summarize our results.

\section{Quasiperiodically forced velocity dependent system}

Let us consider briefly the dynamics of a damped and driven rotating parabola
system and discuss some of its general properties as reported in ref.[11].

\par
A mechanical model describing the motion of a freely sliding particle of unit mass 
on a parabolic wire (z=$\sqrt {\lambda} x^2$) rotating with a constant angular velocity $\Omega$ ($\Omega^2=\Omega_o^2
=-\omega_o^2+g \sqrt(\lambda)$ and
g is the acceleration due to gravity)can be 
associated with a velocity dependent Lagrangian [27]

\begin{equation}
\L={1 \over 2} \left[ (1+ \lambda x^2) \dot x^2 - \omega_o^2 x^2 \right].
\end{equation}

\noindent
Here overdot stands for a derivative with respect to time. The corresponding equation of motion is

\begin{equation}
(1+ \lambda x^2) \ddot x+ \lambda x \dot x^2 +\omega_o^2x = 0,
\end{equation}

\noindent

When $\omega_o^2 > 0$,  Eq.(2) can be integrated in terms of elliptic integrals.
Interesting bifurcations and different routes to chaos occur in the above model
when the system is acted upon by additional damping and external forcing [11].  In
this case, Eq.(2) gets modified to

\begin{equation}
(1+ \lambda x^2) \ddot x+ \lambda x \dot x^2 +\omega_o^2 x+ \alpha \dot x = f \cos \omega t.
\end{equation}

\par
The familiar period doubling bifurcations, preceded by a symmetry breaking
bifurcation, intermittency and antimonotonicity have been identified by us
earlier in ref [11].

\par
Another interesting physical situation is the case in which there is an
additional parametric modulation in the angular velocity,
 $$\Omega = \Omega_o(1+\epsilon \cos \omega_p t), \nonumber$$
so that we can replace $\omega_o^2=g \sqrt{\lambda}-\Omega_o^2$ in Eqs. (1)-(3) by $g \sqrt{\lambda}-\Omega2=\omega_o^2 -\Omega_o^2[2 \epsilon
\cos \omega_p t +0.5 \epsilon^2 (1+ \cos 2 \omega_p t)]$. Then the equation of motion becomes (see for example p.351 in ref. [27] and 
also ref. [11])

\begin{equation}
(1+ \lambda x^2) \ddot x+ \lambda x \dot x^2 +\omega_o^2 x-\Omega_o^2[2 \epsilon
\cos \omega_p t +0.5 \epsilon^2 (1+ \cos 2 \omega_p t)]x+ \alpha \dot x = f \cos \omega_e t,
\end{equation}
where $\epsilon$ is a small parameter.

We have already noted an interesting new quasiperiodic route to chaos, namely
two frequency quasiperiodicity $\rightarrow$
torus doubling $\rightarrow$ torus merging $\rightarrow$ wrinkling $\rightarrow$ strange nonchaotic
attractor $\rightarrow$ chaos in the system (4) in ref.[11]. In this paper 
we make a detailed study of this and 
 many other quasiperiodic routes to chaos  which can occur in this  model in a range of $f$-$\epsilon$ parameter values
and compare them.  For our
 analysis we rewrite the system (4) as
\par

\begin{eqnarray}
\dot x &  = &  y, \nonumber \\
\dot y & = & \left [ {-\lambda x y^2-\{\omega_o^2-\Omega_o^2[2 \epsilon \cos \phi +
0.5 \epsilon^2 (1+ \cos 2 \phi)] \}x - \alpha y + f \cos \theta \over
(1+ \lambda x^2) }  \right], \nonumber    \\
\dot \phi & = & \omega_p,  \nonumber \\
\dot \theta &  = & \omega_e.
\end{eqnarray}

We note that the system (5) remains invariant under the reflection symmetry ($x,y,f$) $\rightarrow$ ($-x,-y,-f$), (or
equivalently (4) under the transformation ($x,f$) $\rightarrow$ ($-x,-f$)). In analogy with low dimensional systems involving 
more than one control parameter when period bubbles occur [28],  one may  expect  remergence 
of torus doubling sequences to occur in this model, which we indeed show to be true in the following.   

\section{Characterization of the  quasiperiodic, strange nonchaotic and chaotic attractors}
There are several quantities to characterize the attractors, which are useful to
distinguish strange nonchaotic from chaotic and quasiperiodic attractors. We briefly review
some of them which we will use in our study.

\subsection{The Lyapunov exponents}

For the system (5), there are two Lyapunov exponents that are trivial in the
sense that they are identically zero by virtue of the two excitation frequencies.
Let the Lyapunov exponents $\lambda_i$ be ordered by their values,$\lambda_1 \geq
\lambda_2 \geq \lambda_3 \geq \lambda_4 $. We then have the following possibilities:
(1) two frequency quasiperiodic attractors, $\lambda_1= \lambda_2=0 > \lambda_3, \lambda_4 $,
(2) three frequency quasiperiodic attractors,  $\lambda_1= \lambda_2= \lambda_3 =0 > \lambda_4 $,
(3) strange chaotic attractors, atleast $\lambda_1 > 0 $ and (4) strange nonchaotic
attractors, the same as two frequency quasiperiodic attractors.

\subsection{Winding number}

The winding number for the orbit of Eq.(5) is defined by the limit

\begin{equation}
W = \lim_{t \rightarrow \infty} \{ {(\alpha(t) - \alpha(t_o)) \over t} \},
\end{equation}

\noindent
where $(x, \dot x) =(r \cos \alpha, r \sin \alpha). $  For the two frequency
quasiperiodic attractors, the winding number satisfies

\begin{equation}
W ={l \over n} \omega_p +{m \over n} \omega_e,
\end{equation}

\noindent
where $l$,$m$, and $n$ are integers.  Combining the winding number and the Lyapunov
exponents, we can distinguish the strange nonchaotic attractors from the other
nonchaotic attractors as noted in Table I.

\subsection{ Power spectrum analysis}

To quantify the changes in the power spectrum (obtained using Fast Fourier Tranform (FFT)
technique) one can compute the so called spectral distribution function $N ( \sigma) $,
defined to be the number of peaks in the Fourier amplitude spectrum larger than
some value say $ \sigma $. Scaling relations have been predicted for $N ( \sigma) $
in the case of two,three frequency quasiperiodic and strange nonchaotic attractors.
These scaling relations are $N(\sigma) \sim \ln {1 \over \sigma}$,$N(\sigma) \sim \ln^2 {\sigma},$
and $N(\sigma) \sim \sigma^{-\beta} $ respectively corresponding  to two, three frequency
quasiperiodic and strange nonchaotic attractors. In the Romeiras and Ott [2]
studies, the power law exponent was found empirically to lie with in the range
$1 < \beta < 2$ for the strange nonchaotic attractor. Thus the above characterizations allow us to
distinguish the strange nonchaotic attractor from other nonchaotic attractors as seen in Table I.

\subsection{ Dimensions}
To quantify geomtric properties of attractors, several methods have been used to
compute the dimension of the attractors. Among them, what we have used is the 
correlation dimension (introduced by Grassberger and Procaccia [26]) which may be computed
from the correlation function C(R) defined as

$$ C(R) =\lim_{N \to \infty} \left [ {1 \over N^2} \sum_{i,j=1}^N H(R- \mid x_i -x_j \mid) \right ], \nonumber $$

\noindent
where $x_i$ and $x_j$ are points on the attractor, H(y) is the Heaviside function
(1 if y $\geq$ 0 and 0 if y $ < $ 0), N is the number of points randomly chosen
from the entire data set. The Heaviside function simply counts the number of points
within the radius R of the point denoted by $x_i$ and C(R) gives the average fraction of 
points. Now the correlation dimension is defined by the variations of C(R) with R:

$$ C(R) \sim R^d    \hskip 20pt as \hskip 20pt  R \rightarrow 0. \nonumber $$
Therefore the correlation dimension (d) is the slope of a graph of log C(R) versus log R.
Once one obtains the  dimensions of the attractors, it will be easy   to quantify strange
property of the attractors as seen in Table I.

\section{Different routes to chaos via  strange nonchaotic attractors}

Now we consider the combined effect of both  the external and parametric forcings in
Eq. (5).  To be concrete, we consider the dynamics of (5) and  numerically integrate it 
using the fourth order Runge-Kutta algorithm with adaptive step size with  the
values of the parameters  fixed at $\omega_o^2$=0.25, $\lambda$=0.5, $\alpha$=0.2,
$\Omega_o^2$=6.7, $\omega_p$=1.0 and $\omega_e$=0.991. Various characteristic quantities such as the
winding numbers, Lyapunov exponents, power spectral measures and dimensions as discussed in the previous section
have been used to distinguish
quasiperiodic, strange nonchaotic and chaotic attractors. Further to identify the different attractors, the dynamical
transitions are traced out by two scanning procedures: (i) varying $f$ at a fixed $\epsilon$ and (ii) varying $\epsilon$
at a fixed $f$. The resulting phase diagram in the ($f$-$\epsilon$) parameter space is shown in Fig.1.  The various
features indicated in the phase diagram are summarised and the dynamical transitions are elucidated in the
following. 

\subsection{Torus breaking bifurcations and the birth of strange nonchaotic attractors}

For low $f$ and low $\epsilon$ values, the system exhibits two frequency quasiperiodic oscillations denoted by 1T in Fig. 1.
When the value of $\epsilon$ exceeds a certain critical value for a fixed low $f$, a transition
from two frequency quasiperiodic (1T) to chaotic attractor(C)  via strange nonchaotic attractor (S) occurs on 
increasing $\epsilon$. For example, we fix the strength of the external forcing parameter value as $f$=0.302 and 
vary the modulation parameter $\epsilon$.
For $\epsilon$=0.03,  Fig. 2a of the attractor has smooth branches and this
indicates that the system is in a two frequency quasiperiodic state.  As $\epsilon$
increases, the branches in the Fig. 2b start to wrinkle (W1).  As
$\epsilon$ increases further, the attractor becomes extremely wrinkled and has
several sharp bends. The sharp bends appear to become actual
discontinuties at $\epsilon$=0.0419  and ultimately result in fractal phenomenon. Such a phenomenon is  essentially the result
of the collision of stable and unstable torus in a dense set of points as was shown by Feudal et al. in their route to
chaos via SNA [17].  At such
values, the nature of the attractor is strange (Fig. 2c) eventhough the largest
Lyapunov exponent in Fig. 3 remains negative.  For this attractor, the correlation dimension is
1.33 while the  Fourier amplitude scaling constant ($\sigma$) is 1.54.  Winding number W 
does not  satisfy the realtion (7) for this attractor.  Hence, these studies confirm further that
the attractor shown in Fig. 2c is strange nonchaotic. As $\epsilon$ increases further, an
attractor visibly similar to Fig. 2c appears (see Fig 2d for $f$=0.042).  However, it has
a positive Lyapunov exponent and hence it corresponds to a chaotic attractor.

\subsection{Remerging torus doubling bifurcations: torus bubble and its consequences}

\subsubsection{Torus bubbling}

On increasing the forcing parameter $f$ further, 0.305$<f<$0.325, the fascinating novel phenomenon of 
torus bubble appears within a range of  
values of $\epsilon$. Within this range of $f$, on increasing the value of $\epsilon$  along the same line, the onset of chaos 
is realised via
strange nonchaotic attractor.
To be more specific, the parameter $f$ is fixed at 0.32 and $\epsilon$ is varied.
For $\epsilon$=0.03, the attractor is a two frequency quasiperiodic attractor (Figs. 4a \& 5a). As 
$\epsilon$ is increased to $\epsilon$=0.0317, the attractor undergoes  a torus
doubling bifurcation (Figs. 4b \& 5b). The corresponding period doubled torus
attractor is denoted as 2T in Fig. 1. We note from Figs. 4 and 5 that the two strands in the
$(x, \phi)$ projection become four strands when torus doubling bifurcation occurs.
When we compute $\phi$ modulo 4$\pi$ instead of 2$\pi$ during integration, we
notice from Figs. 5  that the two bifurcated strands of length 2$\pi$ 
are actually a single strand of length 4$\pi$.  As a result, it can be
concluded that the torus doubling is nonetheless a length doubling bifurcation.
Further, it may noted that this bifurcation is geometrically very similar to that
of period doubling bifurcation in the three dimensional flows.
One then expects as $\epsilon$ is increased further that the doubled attractor has  to continue the doubling sequence as in the case
of period doubling phenomenon. Instead, in the present
case, interestingly the strands of the length doubled attractor begin to merge into that of  a single attractor at $\epsilon$=0.0353
as shown in Fig 6a, leading to the formation of a torus bubble (see Fig. 1), reminiscent of period bubbles in low
dimensional systems [28,29]. On  further increase of the value of $\epsilon$, the transition
from two frequency quasiperiodicity to chaos via strange nonchaotic
attractor takes place due to torus breaking bifurcations as discussed in section 4.A (see Figs 6 b,c,\& d and Fig. 7).

It has been argued in the case of period bubbling in low dimensional systems [28] that the cause of formation of the period bubbles is
essentilally due to the presence of reflection symmetry combined with more than one control parameter present in the system. It appears
that similar argument holds good for the case of higher dimensions for the formation of torus bubbles.

\subsubsection{Formation of multibubbles} 
As the forcing parameter $f$ is increased further in the region 0.325$<f<$0.332, the evolution of attractor undergoes the following transition to 
chaos, wherein more than one bubble is formed on increasing the value of
$\epsilon$: two frequency quasiperiodicity $\rightarrow$ torus doubling $\rightarrow$ wrinkling $\rightarrow$ 
inverse torus doubling (doubled torus) $\rightarrow$ 
merged torus $\rightarrow$ torus bubble $\rightarrow$ merged torus $\rightarrow$ wrinkling $\rightarrow$ SNA $\rightarrow$ chaos. To
illustrate this possibility, let us fix the forcing parameter value as $f$=0.328 and vary the $\epsilon$ value. For $\epsilon$=0.03 the 
attractor is two frequency quasiperiodic. As $\epsilon$ is increased to $\epsilon$=0.0313, the attractor undergoes to torus
doubling bifuraction. The doubled attractor begins to wrinkle when the $\epsilon$ value is increased. However, this wrinkled attractor
appears to become again a torus doubled attractor, instead of approaching the SNA while  the $\epsilon$ value is increased further.  This
 doubled attractor merges into a single torus through inverse bifurcation on increasing the value of $\epsilon$. The merged torus again 
forms a torus bubble and then finally
transits to  chaos via wrinkling and SNA as the value of the $\epsilon$ is incresed further.

On incresing the forcing parameter $f$ further, 0.332$<f<$0.335, the transition from two frequency quasiperiodicity to chaos via 
SNA takes place through the following route, wherein more than two bubbles are formed  as $\epsilon$  increases:two frequency 
quasiperiodicity $\rightarrow$ torus doubling $\rightarrow$ wrinkling $\rightarrow$ inverse torus doubling $\rightarrow$
merged torus $\rightarrow$ torus bubble $\rightarrow$ torus $\rightarrow$ torus bubble $\rightarrow$ torus $\rightarrow$ wrinkling $\rightarrow$ SNA 
$\rightarrow$ chaos. 

\subsection{ Strange nonchaotic and chaotic attractors within and outside the main torus bubble}

On further increase of $f$, $f>$0.335, inside the main torus bubble we observe interesting possiblities of strange nonchaotic and chaotic
attractors  via wrinkling as $\epsilon$ increases. Then two interesting possibilities arise inside the main bubble. The dynamics outside
the main bubble  more or less follows the previous case 3B. The details are as follows.

\subsubsection{ SNA within the main torus bubble}

In a narrow region of $f$, 0.335$<f<$0.339, the SNA undergoes an inverse bifurcation scheme leading to two frequency quasiperiodic attractor 
as $\epsilon$ increases through the following route:
two frequency quasiperiodicity $\rightarrow$ torus doubling $\rightarrow$ wrinkling $\rightarrow$SNA $\rightarrow$  wrinkling
$\rightarrow$ inverse torus doubling (doubled torus) $\rightarrow$ merged torus. For example, the forcing parameter $f$ is fixed at $f$=0.337 and 
$\epsilon $ is varied. For $\epsilon$=0.03 the attractor is a two frequency quasiperiodic one (Fig. 8a ). As $\epsilon$ is increased to $\epsilon$=0.031, 
the attractor undergoes a torus doubling bifuraction ( as seen in Fig. 8b). In lower dimensional systems, the period
doubling occurs in an infinite sequence until the accumulation point is
reached, beyond which chaotic behaviour appears.  However, with tori,in the present case, the truncation
of the torus doubling begins when the two strands become extremely wrinkled (W2)  when 
the $ \epsilon $ value is increased, as shown in Fig. 8c   These
strands  lose  their continuity as well as  smoothness and become strange at $ \epsilon $ =0.0339. At such values, the attractor 
possesses geometrically strange property but does
not obey the sensitivity to the initial conditions (the maximal Lyapunov exponent is negative as seen in Fig. 9) and so it is named as 
strange nonchaotic attractor (Fig. 8d). The emergence of such SNA is 
due to the collision of stable doubled torus and its unstable parent as was shown by Heagy and Hammel[14] in their route. Interestingly 
the SNA (Fig. 8e), instead of approaching a chaotic attrator as the $\epsilon$ value
increases, becomes wrinkled (Fig. 8f) and then torus doubled attractor (Fig. 8g).  The doubled 
attractor again merges into a single torus (Fig. 8h)  on  further increasing the value of $\epsilon$. 

\subsubsection{Chaotic attractor within the main torus bubble}

In a rather large region of $f$, $f>$0.339, the SNA as formed above transits into chaotic attractor on increasing the 
value of $\epsilon$ further through the following route:two frequency quasiperiodicity $\rightarrow$ torus doubling 
$\rightarrow$ wrinkling $\rightarrow$ SNA $\rightarrow$ chaos $\rightarrow$ SNA $\rightarrow$  wrinkling
$\rightarrow$ inverse torus doubling (doubled torus) $\rightarrow$ merged torus. 
To illustrate this possibility, let us
choose the parameter $f$=0.342, and vary the value of $\epsilon$.  For $\epsilon$=0.03
the attractor is a two frequency quasiperiodic attractor (Fig. 10a).  As $\epsilon$
is increased to $\epsilon$=0.0309, the attractor undergoes a torus doubling bifurcation ( Fig. 10b ). 
As $\epsilon$ is increased further the strands of the doubled attractor begin to
wrinkle (W2), as shown in Fig. 10c.
The formation of sharp bends in the strand of the attractor is now clear as $\epsilon$ is increased further. These bends
tend to become actual discontinuities at $\epsilon$= 0.0337, as shown in Fig. 10d. The emergence of  such dicontinuities
on the torus is due to the collision of stable doubled torus and its unstable parent which is similar to the one
found by Heagy and Hammel [14].  At such values, the attractor 
loses smoothness and becomes "strange".  The attractor shown
in Fig. 10d is nothing but strange nonchaotic as the maximum Lyapunov exponent works out to
be $\lambda$=$ -$0.01213 (Fig. 11).  Further the correlation dimension is
1.49, the scaling constant($\sigma$) is 1.38 and winding number  W does not satisfy
the relation (7) for this attractor. Hence, these characteristic studies confirm further that the attractor
shown in Fig. 10d is strange but nonchaotic. On increasing $\epsilon$ value further, for a narrow range of $\epsilon$ values, period doubling bifurcation of SNA is also noticed, which
will be  discussed in detail in the following subsection.  On further increase of the value of  $\epsilon$ to 0.034,
we find the emergence of a chaotic attractor (Fig. 10e), which though visibly similar
to the nonchaotic strange attractor Fig. 10d, has a positive Lyapunov exponent
(see Fig. 11).  The chaotic attractor again becomes SNA when $\epsilon$ is further increased (Fig. 10f). As the value of $\epsilon$ is 
still increased, the SNA becomes  torus doubled attractor (Fig. 10g) via wrinkling. This doubled attractor then merges  into a single
torus (Fig. 10h) 
when the value $\epsilon$ is continuously increased. 

\subsubsection{Dynamics outside the main torus bubble}

Two further interesting transitions  exist  outside of the main torus bubble, namely (i) torus breaking to chaos via SNA and
(ii) torus doubling to chaos via SNA. The details are as follows.

In a narrow region of $f$, 0.335$<f<$0.345, the transition from two frequency quasiperiodicity to chaos via 
SNA takes place outside the main bubble through the following route as $\epsilon$  
increases beyond $\epsilon=$0.0361: torus $\rightarrow$ torus bubble $\rightarrow$ torus $\rightarrow$ torus bubble
$\rightarrow$ torus $\rightarrow$ wrinkling $\rightarrow$ SNA 
$\rightarrow$ chaos. However in the region of $f$, 0.345$<f<$0.352, one also observes a transition from a wrinkled two torus (W2)
to a wrinkled one torus (W1).

Further increase of the value of $f$ beyond 0.352, for $\epsilon$ values greater than 0.0362, introduces yet another kind of transitions
beyond the main bubble as
torus 
$\rightarrow$ torus doubling
$\rightarrow$ wrinkling $\rightarrow$ inverse torus doubling $\rightarrow$ merged torus $\rightarrow$ torus doubling $\rightarrow$ wrinkling 
$\rightarrow$ strange nonchaotic attractor $\rightarrow$ chaos.

\subsection{Period doubling bifurcations of destroyed torus (SNA)  within the main torus bubble}

In the previous subsections, we have seen that the period doubling bifurcation of torus has
been truncated by the destruction of the torus leading to the emergence of strange nonchaotic attractor in certain regions 
of the ($f$-$\epsilon$) parameter space. 
However, we observe in the present system that in some cross sections  of the ($f$-$\epsilon$) parameter space, the period doubling 
bifurcation phenomena still persists in the
destroyed torus, eventhough the actual doubling sequence of the torus has been terminated.
Such a route has also been  recently observed
in coupled Duffing oscillators [8] and in certain maps [22]. The doubling of destroyed torus has been observed in the present model
in a rather long range of $f$,
0.338$<f<$0.358 and for a narrow range of $\epsilon$ values denoted by S2 in Fig.1. 
For example, let us choose $f$=0.345
and vary the value of  $\epsilon$. For $\epsilon$=0.03, the attractor is a two
frequency quasiperiodic torus (Fig. 12a). As $\epsilon$ is increased to $\epsilon$=0.0305,the system
undergoes torus doubling bifurcations (Fig. 12b).  On increase of the value of $\epsilon$ further, 
the attractor begins to wrinkle and finally ends up with fractal nature (SNA) (Fig. 12c). On further increase of the value of $\epsilon$,
the fractal torus undergoes doubling bifurcation (Fig. 12d). If the parameter
$\epsilon$ is still increased, the doubled fractal torus merges into a single fractal torus and finally transits into the  chaotic 
attractor.  

\section{Conclusion}

In this paper, we considered the dynamics of the  specific  example of the quasiperiodically forced
velocity dependent nonpolynomial oscillator system (5) which
illustrates many of  the typical routes to chaos via strange nonchaotic attractors.
It was found that the first two of these routes can be realized in the following 
ways.

\begin{enumerate}
\begin{itemize}
\item {two frequency  quasiperiodicity $\rightarrow $ torus doubling $\rightarrow $ wrinkling
$\rightarrow $ strange nonchaotic attractor $ \rightarrow $ chaos.Here, the birth of SNA is due to the collision
of stable doubled orbit and its unstable parent torus [14] .}

\item{two frequency qusiperiodicity $\rightarrow$wrinkling $\rightarrow$ strange nonchaotic attractor
$\rightarrow$ chaos. In this case, the emergence of SNA is essentially the result of interaction of
stable and unstable torus in a dense set of points [17]. }

\end{itemize}
\end{enumerate}

More interestingly, we have pointed out the novel possibility of the torus bubbling. That is
torus doubling bifurcations in dynamical systems  can under suitable circumstances form finite
sequence which  `merge' in some cross sections of the parameters space, inhibiting
the onset of torus doubling route to chaos. Such remerging bifurcations having
finite number of `bubbles' occur only within some range of the parameters values.An important consequence
is of such remerging is that the orbits become again stable and relatively large regions reappear around them,
where the motion is regular and predictable.
To illustrate such remerging torus
doubling bifurcations,in
our present study, we have shown two  more routes: 

\begin{enumerate}
\begin{itemize}
\item{two frequency quasiperiodicity $\rightarrow $ torus doubling $\rightarrow $
torus merging followed by the gradual fractalization of torus to chaos via SNA}

\item{two frequency quasiperiodicity $\rightarrow $ torus doubling $\rightarrow$ wrinkling
$\rightarrow $ SNA $\rightarrow $ chaos $\rightarrow $ SNA $\rightarrow $ wrinkling $\rightarrow $ inverse torus doubling 
$\rightarrow $ torus $\rightarrow $ torus bubbles followed by the onset of torus breaking to chaos via SNA or 
followed by the onset of torus doubling route to chaos via SNA.}

\end{itemize}
\end{enumerate}
From these routes, it can be concluded that prior to standard routes for transition
to strange nonchaotic
attractor, the possibilities of several bifurcations on the torus can be realised. 

Finally, the period
doubling bifurcations of the destroyed torus have also been observed in our model in a narrow region of the $(f-\epsilon)$
parameter space:
\begin{enumerate}
\begin{itemize}

\item{two frequency quasiperiodicity $\rightarrow$ torus doubling $\rightarrow$ wrinkling $\rightarrow $ destroyed torus $\rightarrow $
period doubling of destroyed torus $\rightarrow$ merged destroyed torus $\rightarrow$ chaos.}

\end{itemize}
\end{enumerate}

\section{Acknowledgements}
This work  forms part of a Department of Science and Technology,
Government of India research project. Also one of the authors (A.V) wishes to
acknowledge the Council of Scientific and Industrial Research, Government of India,
for providing a Senior Research Fellowship.

\newpage
\begin{table}
\caption{Characteristics of attractors}
\begin{tabular}{lcccl}
Types of attractors &  Winding number   & Lyapunov exponents &  Power law relations   & Dimensions \\
&  & & & \\
three frequency & W $\neq {m \over n} \omega_p+{l \over n} \omega_e$ &
$ \lambda_1 < 0, \lambda_2 = \lambda_3 = \lambda_4 = 0$ & $ N(\sigma) = \ln^2 \sigma$ &
 Integers  \\
 quasiperiodic  & & & & \\
&  & & & \\
two frequency & W =$ {m \over n} \omega_p+{l \over n} \omega_e$ &
$ \lambda_1, \lambda_2< 0, \lambda_3 = \lambda_4 = 0$ & $ N(\sigma) = \ln{1 \over \sigma}$ &
 Integers  \\
 quasiperiodic &&&& \\
 &&&&\\
strange nonchaotic & W $\neq {m \over n} \omega_p+{l \over n} \omega_e$ &
$ \lambda_1, \lambda_2< 0, \lambda_3 = \lambda_4 = 0$ & $ N(\sigma) = \sigma^{-\beta}$ &
 Fractals  \\
attractor & & & 1 $  < \beta < $ 2 & \\
\end{tabular}
\label{table1}
\end{table}
\begin{figure}
\caption[]{Phase diagram of the two parameters $(f-\epsilon)$ space exhibited by the system (5). Regions of different attractors are denoted as:
1T - two frequency quasiperiodicity attractor, 2T- torus doubled attractor, W1- wrinkled attractor of period one, W2 - wrinkled attractor
of period two, S - strange nonchaotic attractor,
S2- period doubled strange nonchaotic attractor and C - chaotic attractor.}
\label{Fig.1}
\end{figure}
\begin{figure}
\caption[] {Projection of the two frequency quasiperiodic attractors of Eq. (5)
for $f$=0.302: Poincar\'e plot with $\phi$ mod 2$\pi$ in the (x,$\phi$) plane,
(a) two frequency quasiperiodic attractor at $\epsilon$=0.030,
(b) torus wrinkled attractor for $\epsilon$=0.0405,
(c) strange nonchaotic attractor for $\epsilon$=0.0419,
(d) chaotic attractor for $\epsilon$=0.042,
The other parameters are $\omega_o^2$=0.25,
$\lambda$=0.5, $\alpha$=0.2, $\omega_p$=1.0, $\Omega_o^2$=6.7 and $\omega_e$=0.991.}
\label{Fig.2}
\end{figure}
\begin{figure}
\caption[] { Largest Lyapunov exponent $\lambda_{max}$ vs.$\epsilon$ corresponding to fig. 2}
\label{Fig.3}
\end{figure}
\begin{figure}
\caption[] {Projection of the two frequency quasiperiodic attractors of Eq. (5)
for $f$=0.302: Poincar\'e surface of section in the (x,y) plane (a) torus at $\epsilon$=0.03,
(b)torus doubled attractor at $\epsilon$=0.0317. The other parameters are $\omega_o^2$=0.25,
$\lambda$=0.5, $\alpha$=0.2, $\omega_p$=1.0, $\Omega_o^2$=6.7 and $\omega_e$=0.991.}
\label{Fig.4}
\end{figure}

\begin{figure}
\caption[] {Projection of the two frequency quasiperiodic attractors of Eq.(5)
for $f$=0.302:  Poincar\'e surface of section with $\phi$ mod 2$\pi$ in the (x,$\phi$) plane
(a) two frequency quasiperiodic attractor for $\epsilon$=0.03,
(b) torus doubled attractor at $\epsilon$=0.0317,
(c) \& (d) same as (a) \& (b) except $\phi$ mod 4$\pi$ during integration. The other parameters are $\omega_o^2$=0.25,
$\lambda$=0.5, $\alpha$=0.2, $\omega_p$=1.0, $\Omega_o^2$=6.7 and $\omega_e$=0.991.}
\label{Fig.5}
\end{figure}

\begin{figure}
\caption[] {Projection of the two frequency quasiperiodic attractors of Eq.(5)
for $f$=0.32: Poincar\'e plot with $\phi$ mod 2$\pi$ in the (x,$\phi$) plane,
(a) merged attractor for $\epsilon$=0.0353
(b) wrinkled attractor for $\epsilon$=0.0399
(c) strange nonchaotic attractor for $\epsilon$=0.041
(d) chaotic attractor for $\epsilon$=0.0413
The other parameters are $\omega_o^2$=0.25,
$\lambda$=0.5, $\alpha$=0.2, $\omega_p$=1.0, $\Omega_o^2$=6.7 and $\omega_e$=0.991.}
\label{Fig. 6}
\end{figure}

\begin{figure}
\caption[] { Largest Lyapunov exponent $\lambda_{max}$ vs.$\epsilon$ 
corresponding to Figs. 4,5 \& 6}
\label{Fig. 7}
\end{figure}

\begin{figure}
\caption[] {Projection of the two frequency quasiperiodic attractors of Eq.(5)
for $f$=0.337: Poincar\'e plot with $\phi$ mod 2$\pi$ in the (x,$\phi$) plane,
(a) two frequency quasiperiodic torus at $\epsilon$=0.03,
(b) doubled torus  attractor for $\epsilon$=0.0312,
(c) wrinkled doubled attractor for $\epsilon$=0.0335,
(d) strange nonchaotic attractor for $\epsilon$=0.0342,
(e) strange nonchaotic attractor for $\epsilon$=0.0345,
(f) wrinkled attractor for $\epsilon$=0.03461
(g) doubled torus  attractor for $\epsilon$=0.0347,
(h) merged attractor for $\epsilon$=0.036,
The other parameters are $\omega_o^2$=0.25,
$\lambda$=0.5, $\alpha$=0.2, $\omega_p$=1.0, $\Omega_o^2$=6.7 and $\omega_e$=0.991.}
\label{Fig. 8}
\end{figure}

\begin{figure}
\caption[] { Largest Lyapunov exponent $\lambda_{max}$ vs.$\epsilon$ 
corresponding to Fig.8}
\label{Fig.9}
\end{figure}

\begin{figure}
\caption[] {Projection of the two frequency quasiperiodic attractors of Eq.(5)
for $f$=0.342: Poincar\'e plot with $\phi$ mod 2$\pi$ in the (x,$\phi$) plane,
(a) two frequency quasiperiodic torus at $\epsilon$=0.03,
(b) doubled torus  attractor for $\epsilon$=0.0309,
(c) wrinkled doubled attractor for $\epsilon$=0.033,
(d) strange nonchaotic attractor for $\epsilon$=0.0337,
(e) chaotic attractor for $\epsilon$=0.034,
(f) strange nonchaotic attractor for $\epsilon$=0.0345,
(g) doubled torus  attractor for $\epsilon$=0.0347,
(h) merged attractor for $\epsilon$=0.036,
The other parameters are $\omega_o^2$=0.25,
$\lambda$=0.5, $\alpha$=0.2, $\omega_p$=1.0, $\Omega_o^2$=6.7 and $\omega_e$=0.991.}
\label{Fig. 10}
\end{figure}

\begin{figure}
\caption[] { Largest Lyapunov exponent $\lambda_{max}$ vs.$\epsilon$ 
corresponding to Fig.10}
\label{Fig.11}
\end{figure}

\begin{figure}
\caption[] {Projection of the two frequency quasiperiodic attractors of Eq.(5)
for $f$=0.345: Poincar\'e plot with $\phi$ mod 2$\pi$ in the (x,$\phi$) plane,
(a) two frequency quasiperiodic torus at $\epsilon$=0.03,
(b) doubled torus  attractor for $\epsilon$=0.0305,
(c) strange nonchaotic attractor for $\epsilon$=0.0334,
(d) doubled strange nonchaotic  attractor for $\epsilon$=0.0337,
The other parameters are $\omega_o^2$=0.25,
$\lambda$=0.5, $\alpha$=0.2, $\omega_p$=1.0, $\Omega_o^2$=6.7 and $\omega_e$=0.991.}
\label{Fig. 12}
\end{figure}

\begin{references}
\bibitem{}
C. Grebogi, E. Ott, S.Pelikan and J.A. Yorke, Physica 13D, 261 (1984).
\bibitem{}
F.J. Romeiras and E. Ott, Phys. RevA35, 4404 (1987); F.J. Romeiras, A. Bonderson,
E. Ott, T.M. Andonsen jr. and C. Grebogi, Physica 26D, 277 (1987).
\bibitem{}
Y.C. Lai, Phys. Rev E53, 57 (1996)
\bibitem{}
Y. C. Lai, U. Feudel and C. Grebogi, Phys. Rev E54, 6114 (1996).
\bibitem{}
A. Bonderson, E. Ott and T. M. Antonsen jr., Phys. Rev. Lett 55, 2103 (1985).
\bibitem{}
M. Ding, C.Grebogi and E. Ott, Phys. Rev A39, 2593 (1989); M. Ding and J.A. Scott Relso,
Int. J. Bifurcation and Chaos 4, 553 (1994).
\bibitem{}
J. F. Heagy and W.L. Ditto, J. Nonlinear Sci 1, 423 (1991).
\bibitem{}
J.I. Staglino, J.M. Wersinger, E.E. Slaminka, Physica D92, 164 (1996).
\bibitem{}
T. Yalinkaya and Y. C. Lai, Phys. Rev. Lett. 77, 5040 (1996).
\bibitem{}
T. Kapitaniak and J. Wojewoda, {\it Attractors of Quasiperiodically Forced
Systems}, (World Scientific, Singapore, 1993).
\bibitem{}
A. Venkatesan and M. Lakshmanan, Phys. Rev E55, 4140 (1997).
\bibitem{}
Z. Zhu and  Z. Liu, Int. J. Bifurcation and Chaos 7, 227 (1997).
\bibitem{}
T. Kapitaniak and L. O. Chua, Int. J. Bifurcation and Chaos 7, 423 (1997).
\bibitem{}
J. F. Heagy and S. M. Hammel, Physica 70D, 140 (1994).
\bibitem{}
A. S. Pikovsky and U. Feudal, J. Phys. A27, 5209 (1994).
\bibitem{}
A. S. Pikovsky and U. Feudal, Chaos 5, 253 (1995).
\bibitem{}
U. Feudal, J. Kurths and A.S. Pikovsky, Physica 88D, 176 (1995).
\bibitem{}
S. P. Kuznetsov,  A.S. Pikovsky and U. Feudel, Phys. Rev. E51, R1629 (1995).
\bibitem{}
T. Kapitaniak, Phys. Rev. E47, 1408 (1993).
\bibitem{}
V.S. Anishchensko, T.K. Vadivasova, and O. Sosnovtseva, Phys. Rev. E 53, 4451
(1996).
\bibitem{}
T. Nishikawa and K. Kaneko, Phys. Rev. E54, 6114 (1996).
\bibitem{}
O. Sosnovtseva, U. Feudel, J. Kurths and A. Pikovsky, Phys. Lett A218, 255 (1996).
\bibitem{}
W. L. Ditto, M. L. Spano, H.T. Savage, S.N. Rauseo, J.F. Heagy and E.Ott, Phys.
Rev. Lett 65, 533 (1990).
\bibitem{}
T. Zhou, F. Moss and A. Bulsara Phys. Rev. A45, 5394 (1992).
\bibitem{}
W.X. Ding, H. Deutsch, A. Dingklage and C. Wilke, Phys. Rev. E55, 3769 (1997).
\bibitem{}
P. Grassberger and I. Procaccia, Phys. Rev. Lett. 50, 346 (1983).
\bibitem{}
A. H. Nayfeh and D. T. Mook, {\it Nonlinear Oscillations}, (Wiley Interscience,
New York, 1995).
\bibitem{}
M. Bier and T.C. Bountis, Phys. Lett. A104,239 (1984).
\bibitem{}
M. Lakshmanan and K. Murali, {\it Chaos in Nonlinear Oscillators: Controlling and Synchronization}, (World Scientific,
Singapore, 1996).


\end{references}
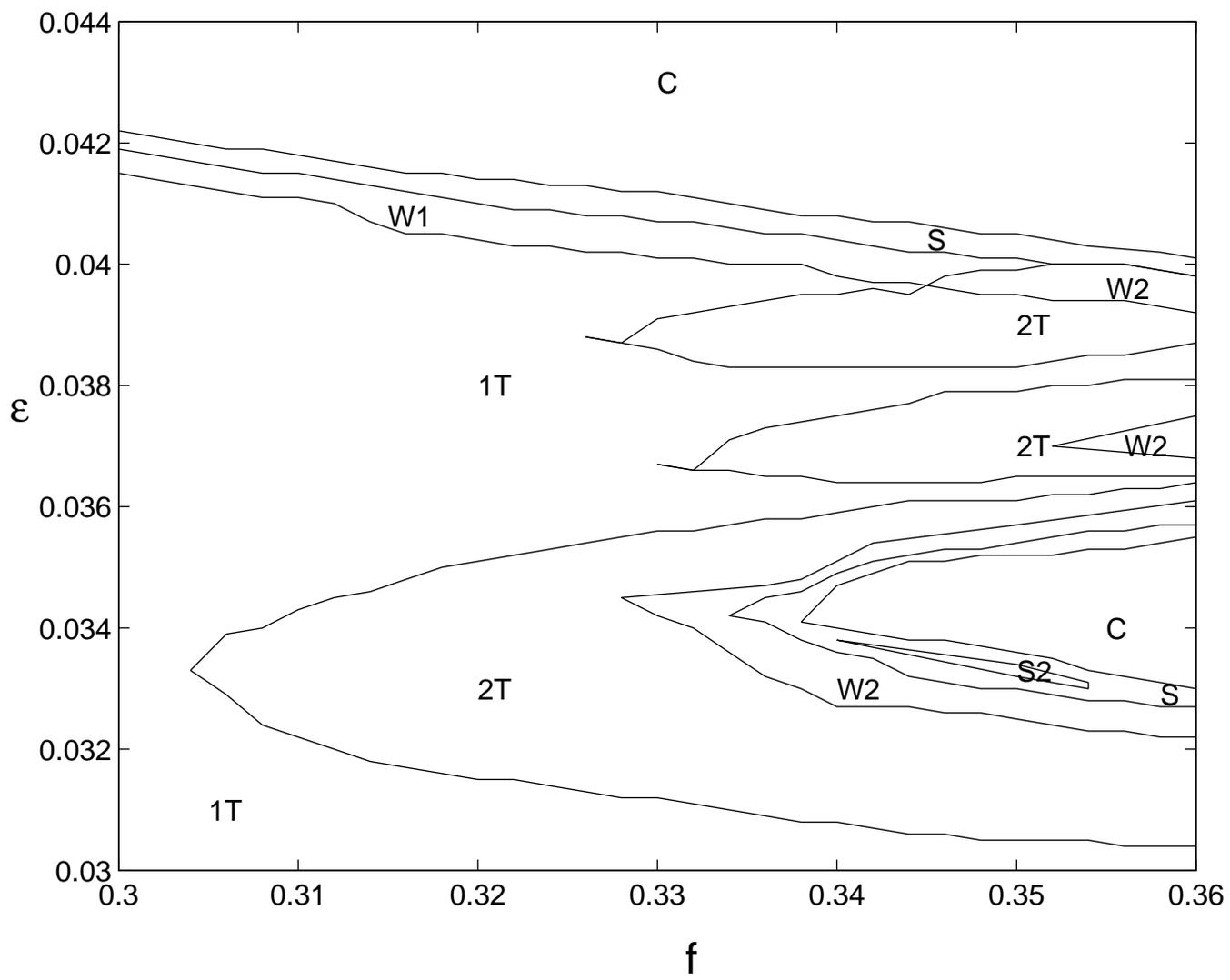
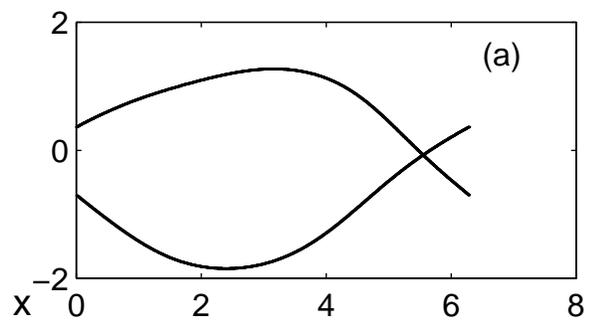
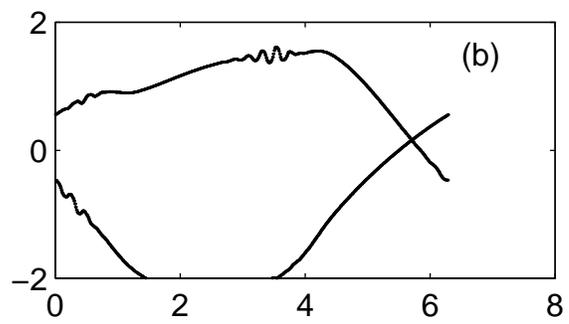
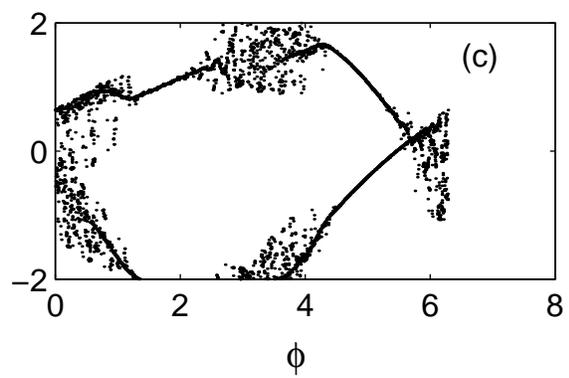
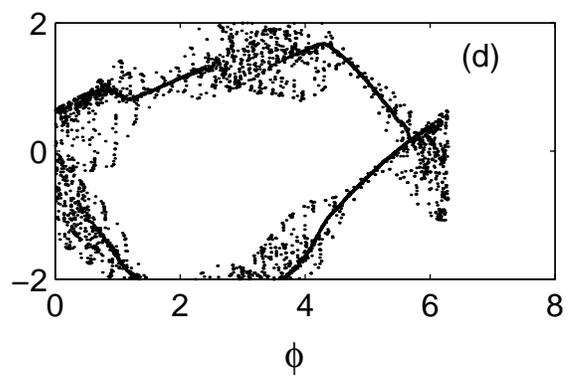
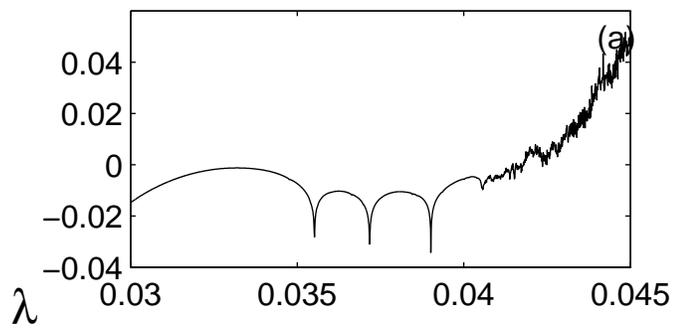
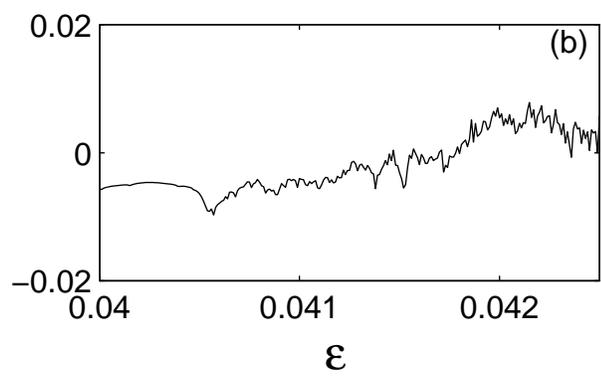
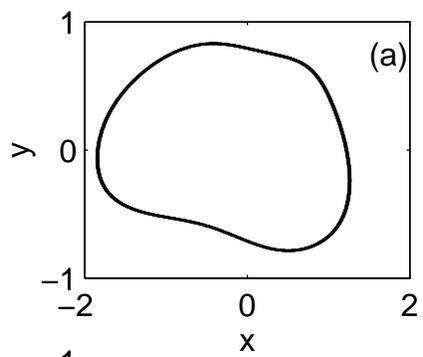
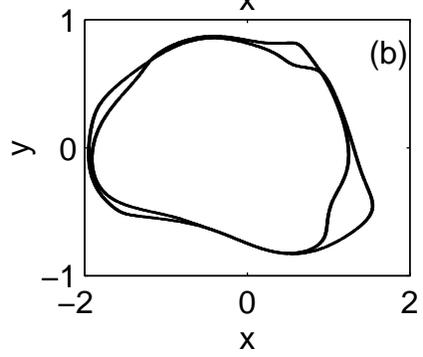
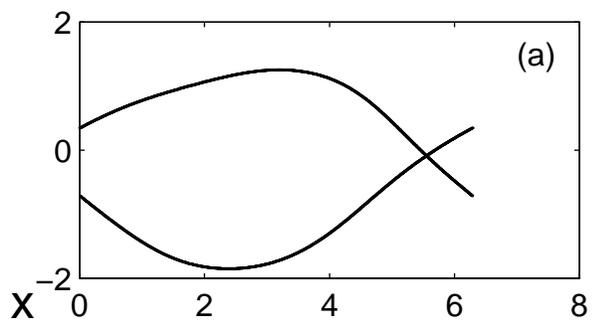
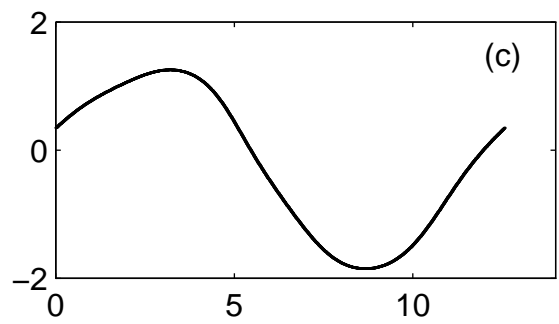
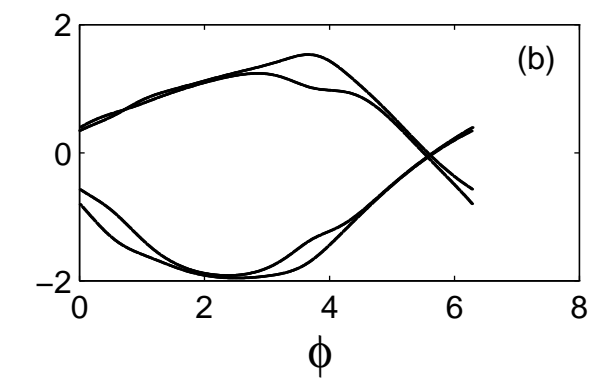
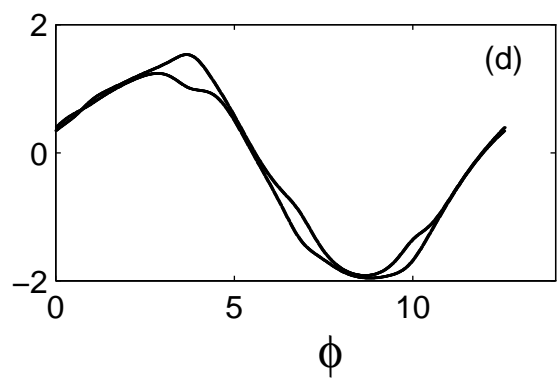
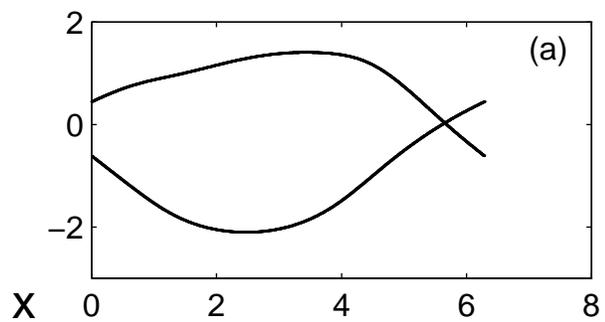
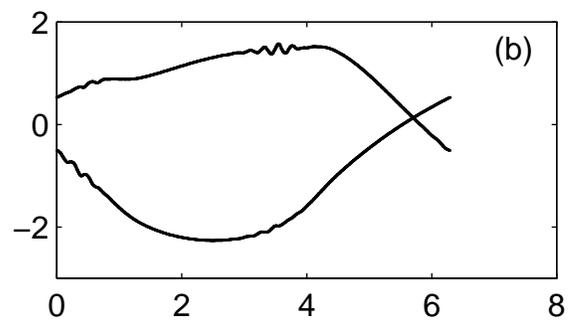
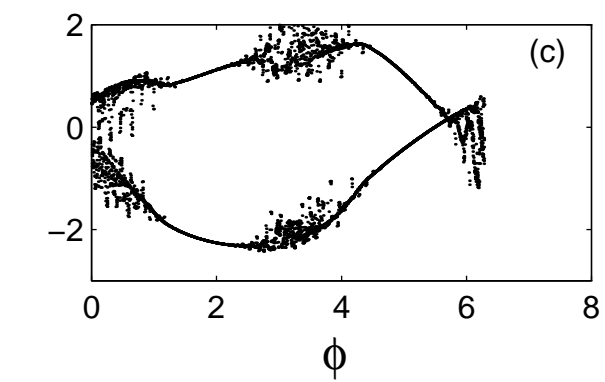
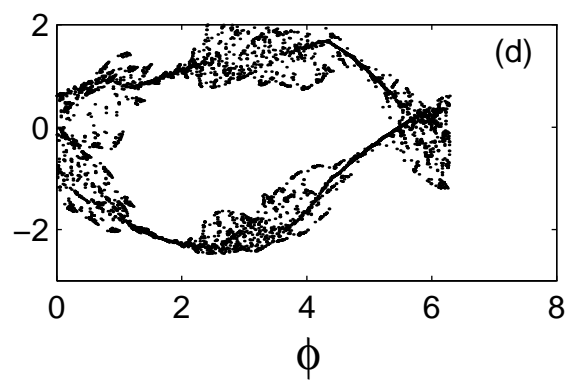
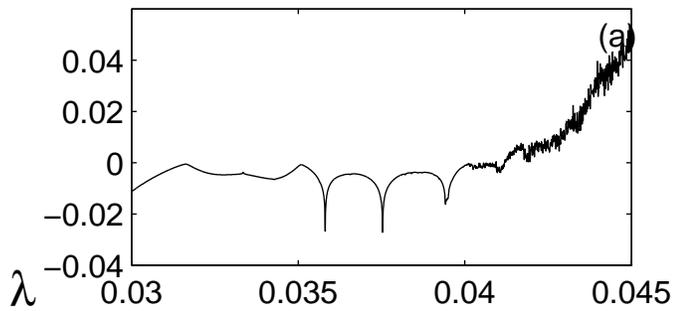
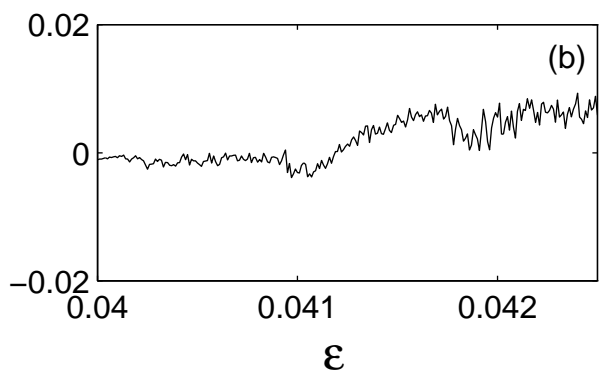
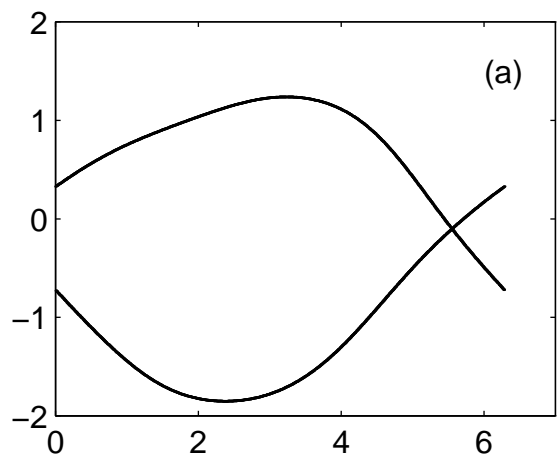
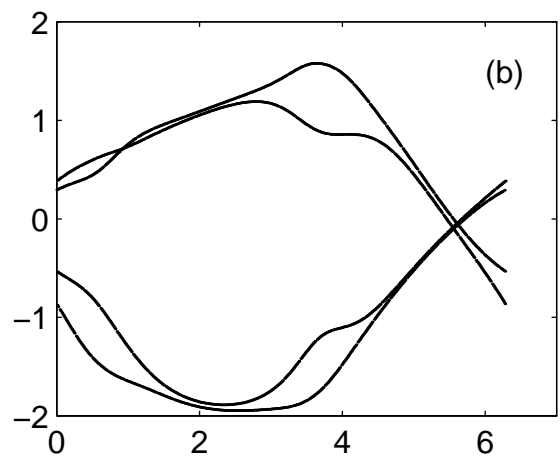
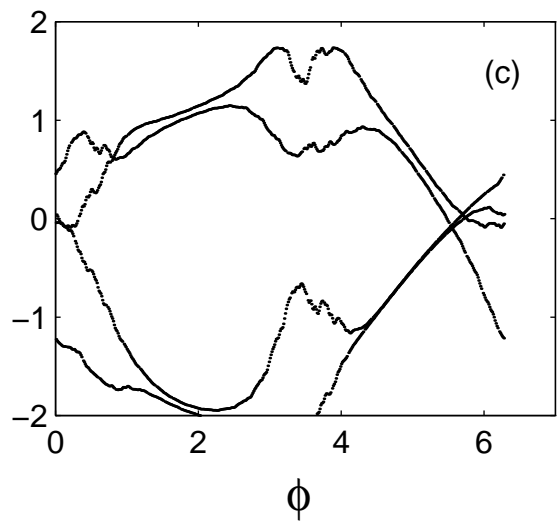
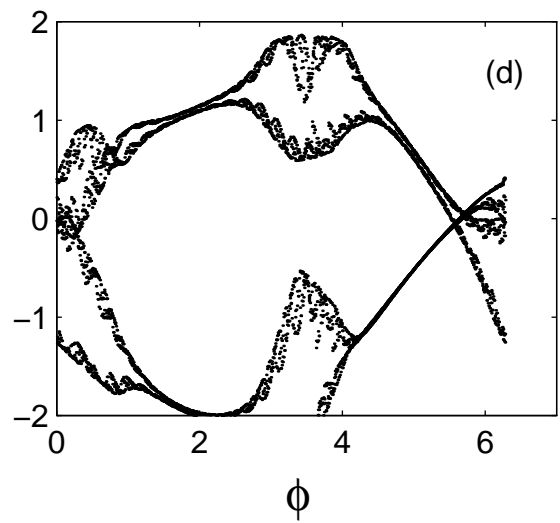
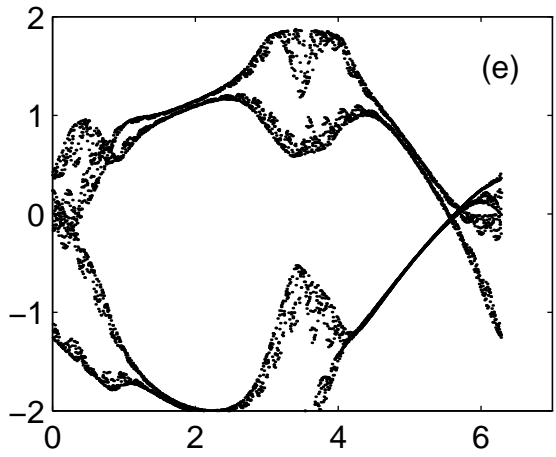
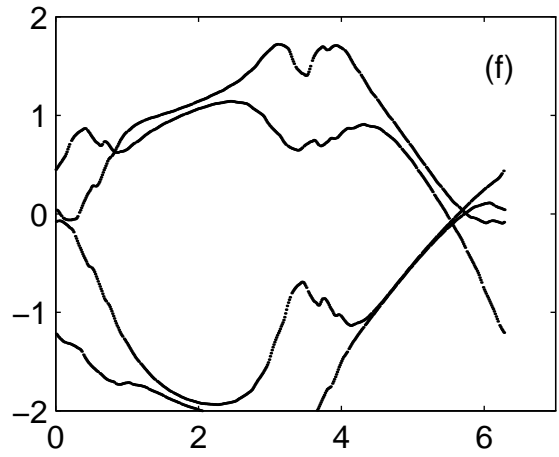
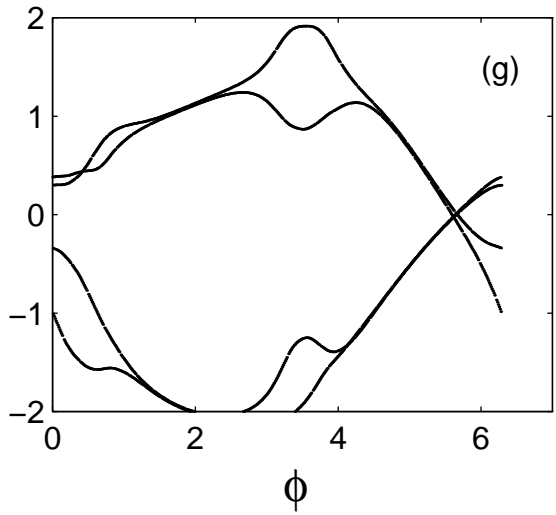
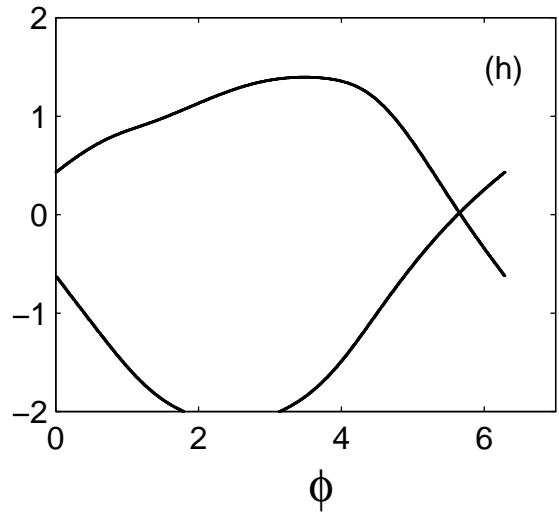
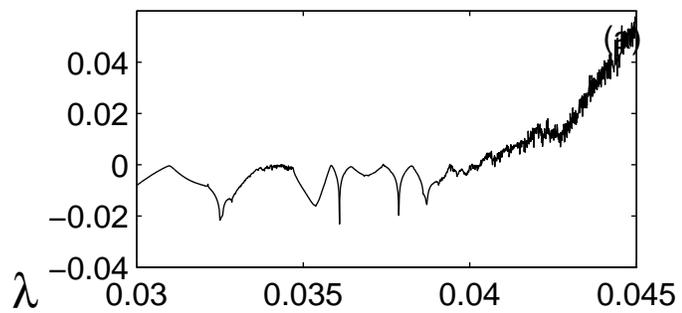
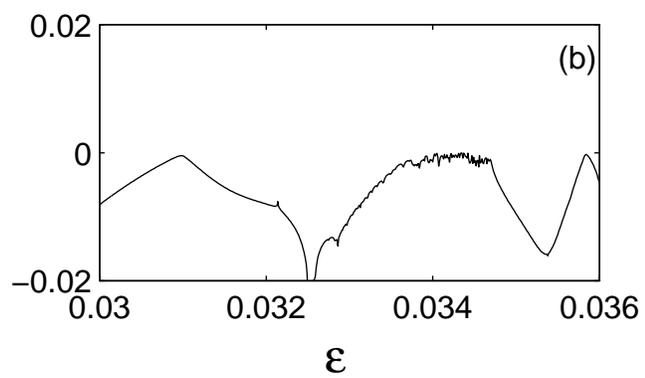
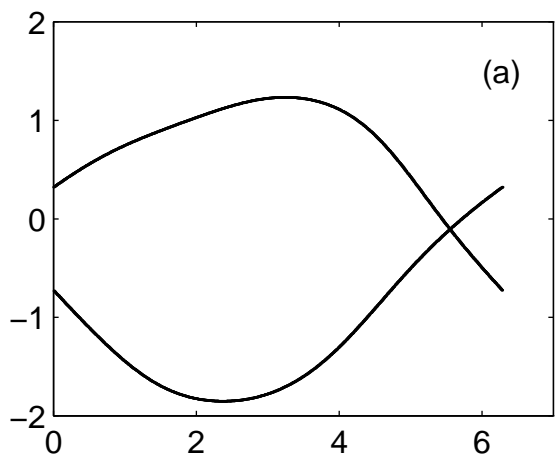
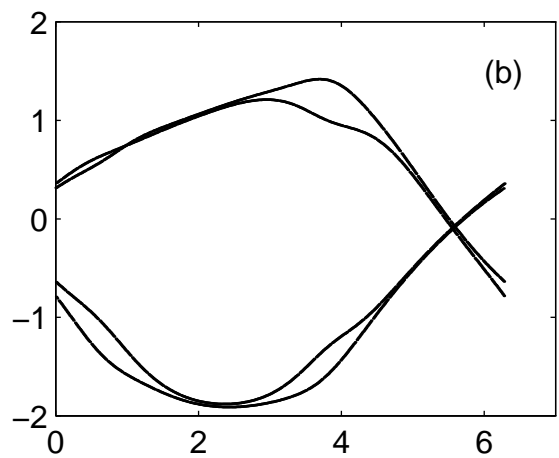
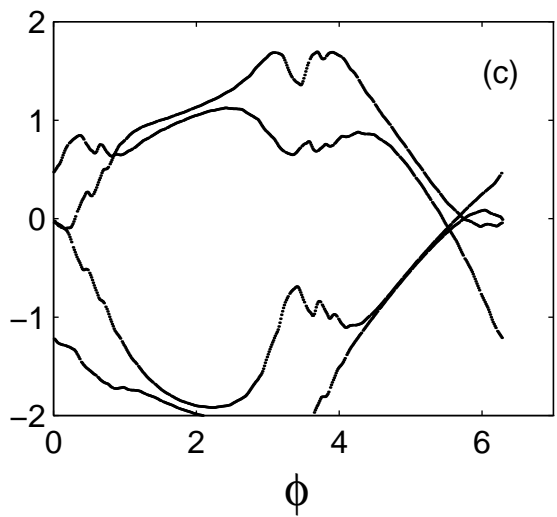
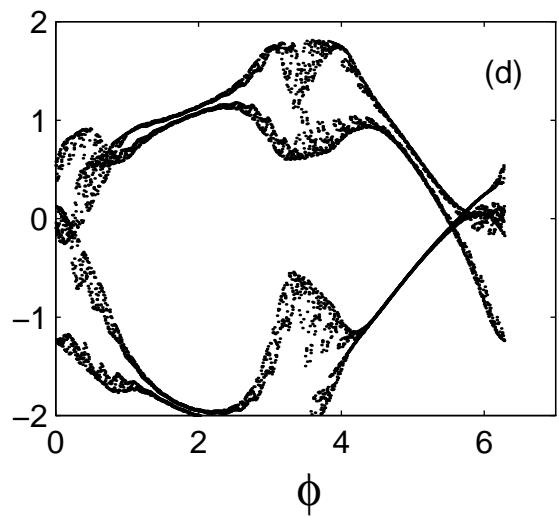
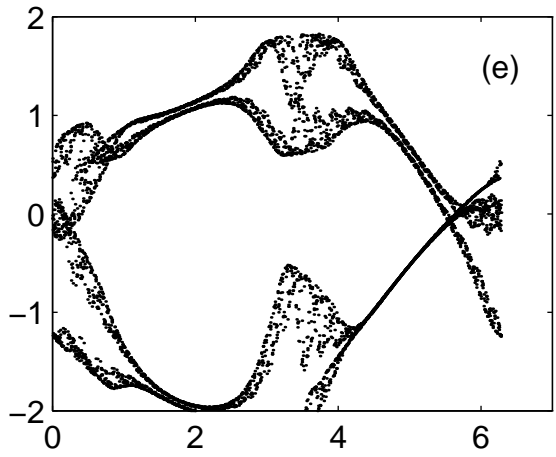
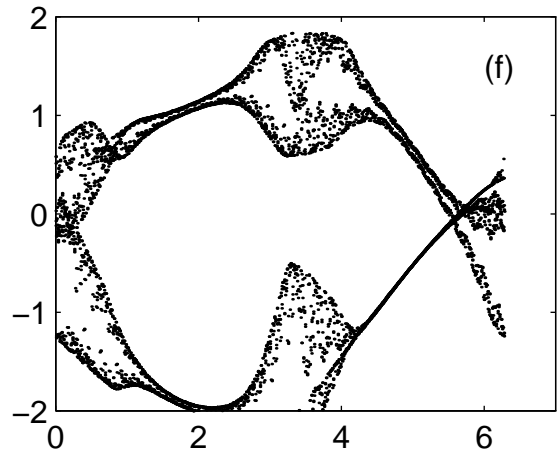
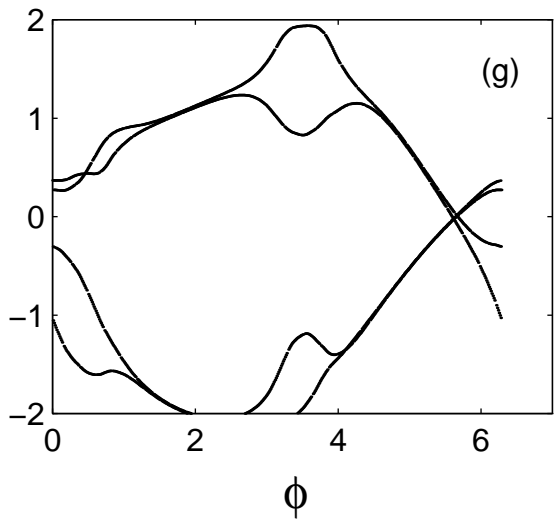
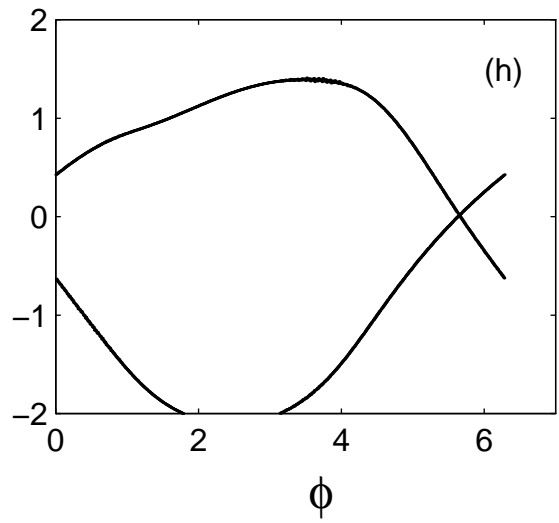
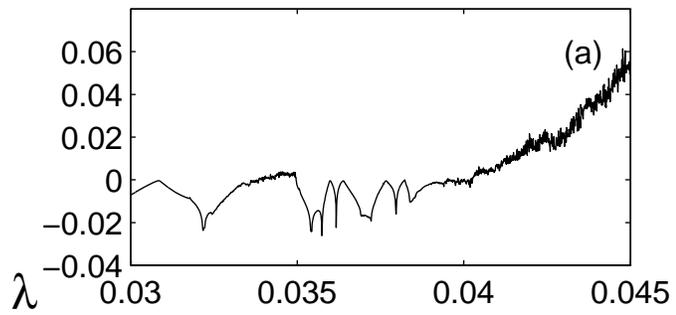
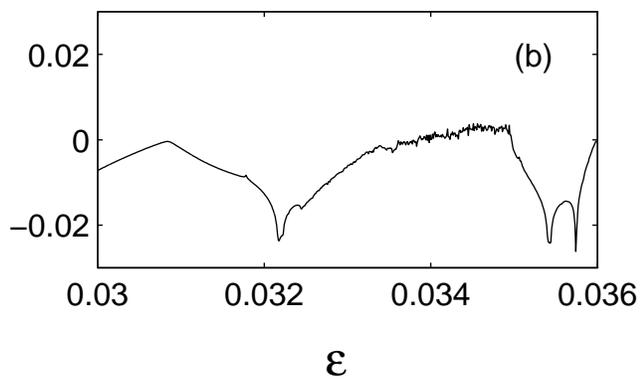
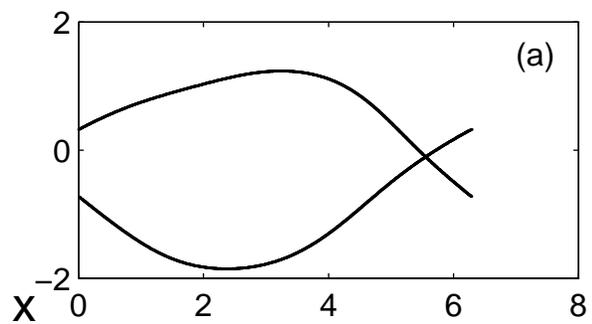
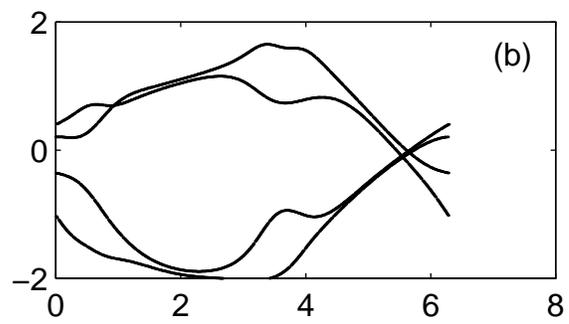
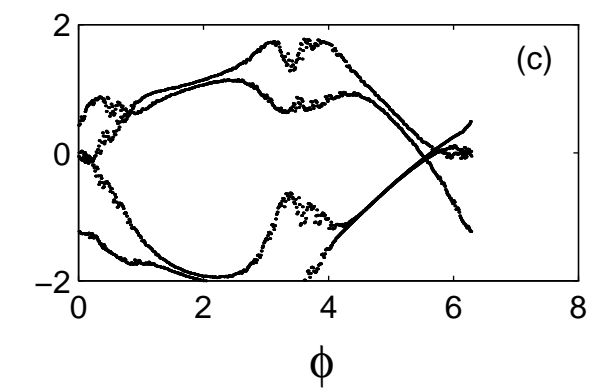
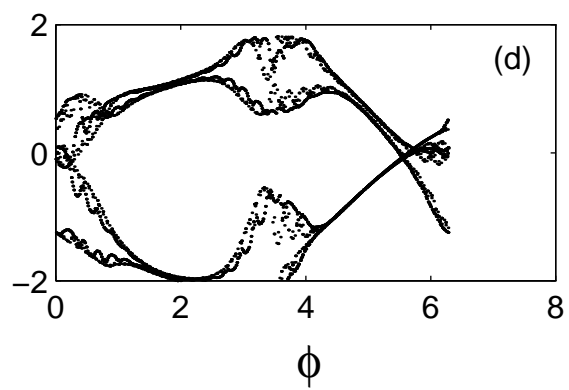
\end{document}